\documentclass[11pt]{article}
\usepackage{latexsym}  
\usepackage{amssymb}
\usepackage{graphicx}
\usepackage{amsmath}

\begin{document}
\hspace*{11cm} {OU-HET-653/2010}

\begin{center}
{\Large\bf Yukawaon Approach to the Sumino Relation}\\
{\Large\bf for Charged Lepton Masses}

\vspace{5mm}
{\bf Yoshio Koide}

{\it Department of Physics, Osaka University,  
Toyonaka, Osaka 560-0043, Japan} \\
{\it E-mail address: koide@het.phys.sci.osaka-u.ac.jp}
\end{center}

\begin{abstract}
On the basis of a supersymmetric yukawaon model,
Sumino's relation for charged lepton masses is re-derived. 
A relation between values of 
$K(\mu) \equiv (m_e +m_\mu + m_\tau)/(\sqrt{m_e} + \sqrt{m_\mu}
+ \sqrt{m_\tau})^2$ and $\kappa(\mu) \equiv \sqrt{m_e m_\mu m_\tau}/
(\sqrt{m_e} + \sqrt{m_\mu}+ \sqrt{m_\tau})^3$ is investigated
without using a relation $K=2/3$. 
Predicted value of $\kappa(\mu)$ is compared with the 
observed value of $\kappa(\mu)$, and it is concluded that
the value $\xi(\mu)\equiv (3/2)K(\mu) -1$ is of the order of 
$10^{-3}$ or less. 
\end{abstract}

\vspace{3mm}

\noindent{\large\bf 1 \ Introduction}

We know an empirical relation \cite{Koidemass}
$$
K \equiv 
\frac{m_e +m_\mu +m_\tau}{(\sqrt{m_e} +\sqrt{m_\mu}
+\sqrt{m_\tau})^2} = \frac{2}{3} .
\eqno(1.1)
$$
In conventional mass matrix models, ``masses" mean not 
``pole masses", but ``running masses".
The relation (1.1) is satisfied with the order of $10^{-5}$ 
for the pole masses, i.e. $K^{pole}=(2/3)\times (0.999989 \pm
0.000014)$, while it is only valid with the order of $10^{-3}$ 
for the running masses.
This has been a mysterious problem as to the relation (1.1) 
for long years.

Recently, a possible solution of this problem has been 
proposed by Sumino \cite{Sumino09PLB,Sumino09JHEP}:
He considers that a flavor symmetry is gauged, and 
radiative corrections by photon to the charged 
lepton masses are exactly canceled by those by  
flavor gauge bosons.
Therefore, as far as the charged lepton masses are
concerned, the running mass values are exactly equal to 
the pole mass values. 
(Hereafter, we will refer to this mechanism as Sumino's mechanism.)
Moreover, as a byproduct of his model, he has obtained
a charged lepton mass relation \cite{Sumino09JHEP}
$$
m_\mu^{3/2} +\sqrt{m_e m_\mu m_\tau} = \sqrt{ m_e m_\tau}
(\sqrt{m_e}+\sqrt{m_\tau}) .
\eqno(1.2)
$$
(We also refer to the relation (1.2) as Sumino's 
relation on the charged lepton masses.)
Then, we can completely determine the charged lepton 
mass spectrum by using two relations (1.1) and (1.2)
simultaneously.
As we demonstrate it later on, a predicted value of 
the electron mass $m_e$ is in fairly good agreement 
with the observed one. 
The Sumino model is an effective theory with 
an energy scale $\Lambda \sim 10^2$ TeV, so that 
many visible effects are fruitfully expected in TeV 
region physics. 
Although the topic investigated by Sumino appears to be 
a narrow and restricted topic, we think that the problems
proposed by Sumino will provide 
a vital clue to new physics.

In the Sumino model, an effective Yukawa coupling 
constant $Y_e^{eff}$ in the charged lepton sector is 
given by $Y_e^{eff} \propto \langle \Phi \rangle 
\langle \Phi \rangle^T$, where $\langle \Phi \rangle$ is 
a vacuum expectation value (VEV) of a scalar field
$\Phi$ with $3\times 3$ components.
Therefore, the quantity $K$ defined in Eq.(1.1) is
expressed by 
$$
K=\frac{v_1^2+v_2^2+v_3^2}{(v_1+v_2+v_3)^2},
\eqno(1.3)
$$
where $(v_1, v_2, v_3)$ are eigenvalues of 
$\langle \Phi \rangle$.
A derivation of the relation $K=2/3$ from
a scalar potential model was first tried by the author
\cite{K-mass90}, where $Y_e^{eff} \propto \langle \Phi \rangle 
\langle \Phi \rangle$ was also assumed. 
However, in Ref.\cite{K-mass90}, the field $\Phi$ was assigned 
to ${\bf 8}+{\bf 1}$ of a U(3) flavor symmetry, 
so that $Y_e \propto \Phi\Phi$ was also assigned to 
${\bf 8}+{\bf 1}$ of U(3), 
while, in the Sumino model, $\Phi \Phi^T$ is assigned to 
${\bf 6}$ of U(3).
In other words, Sumino has assigned the charged lepton 
fields $e_{L}=(e_1, e_2, e_3)_L$ and $e_{R}=(e_1, e_2, e_3)_R$ 
to ${\bf 3}$ and ${\bf 3}^*$ of U(3), respectively.
This assignment is essential for the cancellation between
interactions of photon and flavor gauge bosons.
  
In the derivation of the relation (1.2), 
Sumino \cite{Sumino09JHEP} has proposed a scalar potential for $\Phi$
$$
V=\lambda_1^S (v_1^2 v_3^2 + v_2^2 v_2^2 + v_3^2 v_1^2)
+\lambda_2^S (v_1^4+v_2^4+v_3^4) ,
\eqno(1.4)
$$
where, for convenience, we have expressed his potential
in terms of the eigenvalues of $(v_1, v_2, v_3)$. 
Under the subsidiary condition 
$(v_1^2+v_2^2+v_3^2)/(v_1+v_2+v_3)^2=2/3$, 
minimizing conditions of 
the potential (1.4) lead to a VEV relation
$$
v_2^3+ v_1 v_2 v_3 - v_1 v_3 (v_1+v_3)=0 ,
\eqno(1.5)
$$
in the limit of $\lambda_2^S/\lambda_1^S \rightarrow 0$.
The VEV relation (1.5) means the charged lepton mass 
relation (1.2).
(Note that the relation (1.1) is totally symmetric under 
permutations among $m_e$, $m_\mu$ and $m_\tau$, while the 
relation (1.2) is symmetric only under 
$m_e \leftrightarrow m_\tau$.)
Then, by using the relations (1.1) and (1.2) simultaneously, 
we can completely determine the charged lepton mass spectrum.
It is convenient to express  the charged lepton
mass spectrum in terms of the following two quantities, 
$K$ and $\kappa$ \cite{e-spec-PLB09}, where $K$ is 
defined by Eq.(1.1) and $\kappa$ is defined by
$$
\kappa \equiv \frac{\sqrt{m_e m_\mu m_\tau}}{
(\sqrt{m_e} +\sqrt{m_\mu} +\sqrt{m_\tau} )^3}
=\frac{v_1 v_2 v_3}{(v_1+v_2+v_3)^3}  ,
\eqno(1.6)
$$
which is sensitive to the value of $m_e$.
By using Eqs.(1.1) and (1.2), we obtain 
a predicted value $\kappa =2.20869 \times 10^{-3}$, 
which is in fairly 
good agreement with the observed value 
$$
\kappa^{pole}=(2.0633 \pm 0.0001) \times 10^{-3} .
\eqno(1.7)
$$
Of course, we can exactly fit a predicted value of 
$\kappa$ in the Sumino model to the observed value 
$\kappa^{pole}$ by adjusting the parameter
$\lambda_2^S/\lambda_1^S$.
However, we would like to notice that the relation (1.2) 
has already given an almost reasonable value of 
$m_e$ without help of the second term ($\lambda_2$-term). 

We have a great interest in the Sumino model. 
In this paper, we try to derive the Sumino relation (1.2) 
on the basis of another approach, i.e. on the basis of the 
so-called supersymmetric 
yukawaon model \cite{yukawaon,e-yukawaon-PRD09}, where the effective 
coupling constants $Y_f^{eff}$ ($f=u, d, \nu, e$) in the Yukawa 
interactions are given by $Y_f^{eff}= \langle Y_f \rangle/\Lambda$
(we call the fields $Y_f$ ``yukawaons"), 
and the VEV of $Y_e$ is also related to a field $\Phi$
as $\langle Y_e \rangle \propto \langle\Phi\rangle
\langle\Phi\rangle$.
In the yukawaon model, VEV relations are derived from 
supersymmetric (SUSY) vacuum conditions, so that such
relations are only valid at a high energy scale $\mu \sim \Lambda$.
Since the flavor symmetry is completely broken at 
$\mu \sim \Lambda$, so that the effective coupling constants
$Y_f^{eff}$ evolve as those in the standard SUSY model below 
the scale $\Lambda$.
In the yukawaon model, we consider that the value of 
$K(\mu)$ is not exactly 2/3, but, rather, the value may 
be somewhat deviated from $K=2/3$, i.e. 
$$
K(\mu) =\frac{2}{3} ( 1+ \xi(\mu)),
\eqno(1.8)
$$ 
with $\xi \sim 10^{-3}$ at $\mu \sim \Lambda$. 
Similarly, in the yukawaon model, the value of $\kappa(\Lambda)$ 
does not need to be the observed value (1.7), 
so that we can regard the predicted value $\kappa=2.209\times 10^{-3}$ 
from the Sumino relation (1.2) as a value of $\kappa(\mu)$ at 
$\mu \sim \Lambda$ in the yukawaon model.
For a reference, we list values of $\xi(\mu)$ and $\kappa(\mu)$
in Table 1.
The running mass values used in Table 1 have been quoted from 
Ref.\cite{Xing08}.
Also, in Fig.1, we plot those values as a $\xi$-$\kappa$ curve.
The data points from the left to the right in Fig.1 correspond to 
$\xi$-$\kappa$ values at $\mu=m_Z$, $10^3$ GeV, $10^9$ GeV,
$10^{12}$ GeV and $2\times 10^{16}$ GeV, respectively. 
The ``observed" $\xi$-$\kappa$ curve is dependent 
on the value of $\tan\beta$ in the estimates of the running 
charged lepton masses. 
(Hereafter, we refer to the values of $\xi(\mu)$ and $\kappa(\mu)$
which are calculated by the running mass values $m_{ei}(\mu)$ 
estimated from the observed (pole) masses $m_{ei}$ as  
``observed" values of  $\xi(\mu)$ and $\kappa(\mu)$ in contrast
to the ``predicted" values of  $\xi(\mu)$ and $\kappa(\mu)$ 
in the present model.) 

As in seen in in Fig.1, it is worthwhile noticing
that both $\xi$-$\kappa$ curves for $\tan\beta=10$ and 
$\tan\beta=50$ are almost plotted on a common linear curve. 
We can confirm that the deviation from the common linear 
relation is of the order of $10^{-2}$ or less.
Therefore, the relation given in Fig.1 is useful to 
compare our prediction values of  $\xi(\mu)$-$\kappa(\mu)$ 
with the observed values without assuming a specific value
of $\tan\beta$.

\begin{table}
\caption{Energy scale dependence of $\xi(\mu)$ and $\kappa(\mu)$ 
in cases with $\tan\beta=10$ and $\tan\beta=50$.  
$\xi$ and $\kappa$ for pole masses are given by 
$\xi^{pole}=(0.011\pm 0.014)\times 10^{-3}$ and
$\kappa^{pole}=(2.0633\pm 0.0001)\times 10^{-3}$.
The running mass values of the charged leptons have been quoted from 
Ref.\cite{Xing08}.}

\vspace{2mm}

\begin{tabular}{|c|c|c|c|c|} \hline
       & \multicolumn{2}{c|}{$\tan\beta=10$} &
\multicolumn{2}{c|}{$\tan\beta=50$} \\ \hline       
Scale & $\xi(\mu)$ [$10^{-3}$] & $\kappa(\mu)$ [$10^{-3}$] &
$\xi(\mu)$ [$10^{-3}$] & $\kappa(\mu)$ [$10^{-3}$] \\ \hline
$\mu = m_Z$ & $1.879 \pm +0.002$ &  $2.0276 \pm 0.0002$ &
$1.879 \pm +0.002$ &  $2.0276 \pm 0.0002$ \\
$\mu = 10^3$ GeV & $1.95 \pm 0.02$ & $2.0271\pm 0.0002$ &
$ 2.50\pm 0.02$ & $2.0219 \pm 0.0002$ \\
$\mu = 10^9$ GeV & $2.20 \pm 0.02$ & $2.0247\pm 0.0002$ &
$5.69 \pm 0.02$ & $1.9924 \pm 0.0002$ \\
$\mu = 10^{12}$ GeV & $2.30 \pm 0.02$ & $2.0238 \pm 0.0002$ &
$7.03\pm 0.02$ & $1.9800 \pm 0.0002$ \\
$\mu =2\times 10^{16}$ GeV & $2.42 \pm 0.02$ & $2.0227 \pm 0.0002$ &
$8.62\pm 0.02$ & $1.9654 \pm 0.0002$ \\ \hline
\end{tabular}
\end{table}

In the Sumino model, the subsidiary condition 
$(v_1^2+v_2^2+v_3^2)/(v_1+v_2+v_3)^2=2/3$ [i.e. the relation (1.1)]
was indispensable in deriving the Sumino relation (1.5).
In the present paper, we try to derive the Sumino relation (1.2) 
[or (1.5)] without assuming the relation (1.1), i.e. we derive the 
Sumino relation which is valid even when the relation (1.1) 
is not satisfied.
Although, for the charged lepton mass spectra, some of ideas 
\cite{Koide-U3-PLB08,e-yukawaon-PRD09,e-spec-PLB09} have 
already been proposed on the basis of a yukawaon model, we do not
adopt such a model in the present paper.
Therefore, the present yukawaon model does not predict a value of 
$K(\Lambda)=(2/3)(1+\xi(\Lambda))$, but we can give only a relation 
between values $\xi(\mu)$ and $\kappa(\mu)$ numerically.
We will speculate a possible value $\xi(\Lambda)$  
from the observed value of $\kappa(\Lambda)$. 

In Sec.2, we give a supersymmetric yukawaon model in the 
charged lepton sector.  
A VEV form $\langle\Phi\rangle$ in the present model takes
a diagonal form $\langle\Phi\rangle = {\rm diag}(v_1, v_2, v_3)$ 
in contrast to $\langle\Phi\rangle$ in the Sumino model 
in which $\langle\Phi\rangle$ cannot be diagonal although 
$\langle\Phi\rangle \langle\Phi\rangle^T$ is diagonal.
Therefore, the present model is very intuitive, but it 
needs many additional fields concerned with mass relations,
so that it will somewhat be complicated rather than the
Sumino model.
The present model will contain two parameters.
One is $\xi(\mu)$ defined in Eq.(1.8), because we do not 
consider $K=2/3$.
Another one is a parameter $\rho$ which is given by a ratio of 
coefficients of the superpotential terms.
In the limit of $\xi \rightarrow 0$ and $\rho \rightarrow 0$, 
the result becomes the Sumino mass relation (1.2).
In Sec.3, in order to compare our results to the observed 
values given in Table 1, we will speculate that the value 
of $\rho$ is given by $\rho=3/2$.
Then, we can completely determine the value of $\kappa(\mu)$ 
as a function of $\xi(\mu)$.
We will find that two options are possible:
one is to accept the Sumino mechanism, because a  
predicted value of $\kappa$ at $\xi=0$ is in favor of 
the observed value of $\kappa^{pole}$, and another one
is to consider that the energy scale $\Lambda$ of the effective
theory is $\Lambda \sim 10^4$ GeV, because our result shows
$\kappa(\mu)^{pred} = \kappa(\mu)^{obs}$ at $\mu \sim 10^4$
GeV.
In Sec.2, we assume ad hoc forms of VEV matrices 
for some specific fields, for example, $\langle E \rangle
= v_E {\bf 1}$, and so on.
In Sec.4, we will discuss a possible scenario which provides 
such specific forms.
Finally, Sec.5 is devoted to conclusions and remarks.

\vspace{3mm}

\noindent{\large\bf 2 \ Model}

Stimulated by the Sumino model, in this paper, we assign 
the yukawaon $Y_e$ to ${\bf 6}$ of U(3), not to ${\bf 8}+{\bf 1}$
of U(3) \cite{Koide-U3-PLB08} (and also not to ${\bf 5}+{\bf 1}$ 
of O(3) \cite{Koide-O3-PLB08}), so that the would-be
Yukawa interaction in the charged lepton sector is given by
$$
H_{Yukawa}= \frac{y_e}{\Lambda} \sum_{i,j} \ell^i (Y_e)_{ij}
(e^c)^j H_d ,
\eqno(2.1)
$$
where $\ell=(\nu_L, e_L)$ and $e^c$ are SU(2)$_L$ doublet 
and singlet fields of the leptons, respectively, and $H_d$ is
the down-type SU(2)$_L$ doublet Higgs scalar in the 
conventional model. 
We list fields which are concerned with the charged 
lepton masses in Table 2. 
In order to distinguish each yukawaon $Y_f$ from others, we 
assume a U(1)$_X$ symmetry, and we assign a U(1)$_X$ charge
(``sector" charge) $Q_X$ as $Q_X(Y_f)= x_f$, $Q_X(f^c)=-x_f$, 
and $Q_X=0$ for all SU(2)$_L$ doublet fields.
However, in this paper, since we deal with only fields concerned 
with $Y_e$, we denote $Q_X(Y_e)=x_e$ as $Q_X(Y_e)=+1$ simply.
(Also, hereafter, we denote $Y_e$ as $Y$ simply.)
In the $R$ charge assignments in Table 2, the values of $r_1$ 
and $r_2$ must be chosen as the charges of $Y$, $E$ and $\Theta_T$ 
(and also those of $\bar{Y}$, $\bar{E}$ and $\bar{T}$) are 
different among them, and as an unwelcome combination 
$E\bar{T} E\bar{T}$ is forbidden.
At present, we do not specify those values $r_1$ and $r_2$.
The Z$_2$ parity in the Table 2 has been introduced for the 
purpose to forbidden a term which includes $\bar{T}$ of
an odd number such as ${\rm Tr}[\Phi \tilde{\Phi} E \bar{T}]$. 

\begin{table}
\caption{Quantum numbers of fields concerned with the charged 
lepton masses}

\begin{center}
\begin{tabular}{|c|cccccccc|} \hline
Field & $Y$ & $E$ & $\Theta_T$ & $\bar{Y}$ & $\bar{E}$ & 
$\bar{T}$ & $\Phi$ & $\tilde{\Phi}$ \\ \hline
U(3) & ${\bf 6}$ & ${\bf 6}$ & ${\bf 6}$ & ${\bf 6}^*$ &
${\bf 6}^*$ & ${\bf 6}^*$ & ${\bf 8}+{\bf 1}$ &
${\bf 8}+{\bf 1}$ \\
$Q_X$ & $1$ & $1$ & $1$ & $-1$ & $-1$ & $-1$ & $0$ & $0$ \\
$Q_R$ & $0$ & $r_1$ & $r_1+2 r_2$ & $2-r_1-2r_2$ & $2r_2$ & 
$1-\frac{1}{2}r_1 -r_2$ & $r_2$ & $1-r_2$ \\
Z$_2$ & $+$ & $+$ & $+$ & $+$ & $+$ & $-$ & $+$ &
$+$ \\ \hline
\end{tabular}
\end{center}
\end{table}

According to the quantum numbers given in Table 2, a possible
superpotential $W$ is given as follows: $W=W_Y +W_T +W_S$,
$$
W_T =\mu_T {\rm Tr}[\Theta_T \bar{Y}] + \frac{\lambda_T}{\Lambda}
{\rm Tr}[\Theta_T \bar{T} Y \bar{T}],
\eqno(2.2)
$$
$$
W_Y =\frac{\lambda_E}{\Lambda}{\rm Tr}[Y \bar{E} E \bar{Y}]
+ \frac{\lambda_Y}{\Lambda}{\rm Tr}[\tilde{\Phi}\tilde{\Phi}
Y \bar{E}] + \frac{\bar{\lambda}_Y}{\Lambda} {\rm Tr}[{\Phi}
{\Phi} E \bar{Y}] ,
\eqno(2.3)
$$
$$
W_S = \frac{\lambda_1}{\Lambda}{\rm Tr}[\Phi\Phi\tilde{\Phi}\tilde{\Phi}]
+\frac{\lambda'_1}{\Lambda}{\rm Tr}[\Phi\tilde{\Phi}\Phi\tilde{\Phi}]
+ \frac{\lambda_2}{\Lambda}{\rm Tr}[\Phi\Phi]
{\rm Tr}[\tilde{\Phi}\tilde{\Phi}]
+ \frac{\lambda'_2}{\Lambda}{\rm Tr}[\Phi\tilde{\Phi}]
{\rm Tr}[\Phi\tilde{\Phi}]  
+ \frac{\lambda_3}{\Lambda} {\rm Tr}^2[\Phi] 
{\rm Tr}^2[\tilde{\Phi}] .
\eqno(2.4)
$$
Here, the potential $W_T$ plays a role in giving a relation between
$Y$ and $\bar{Y}$.
The potential $W_Y$ plays a role in providing VEV relations 
$Y \propto \Phi E \Phi^T$ and so on.
Since $\Phi$ and $\tilde{\Phi}$ are ${\bf 8}+{\bf 1}$ of 
U(3), we can consider terms  ${\rm Tr}[\tilde{\Phi} Y \tilde{\Phi}^T
\bar{E}]$ and ${\rm Tr}[Y\tilde{\Phi}^T \tilde{\Phi}^T \bar{E}]$
in addition to the term  ${\rm Tr}[\tilde{\Phi}\tilde{\Phi}
Y \bar{E}]$. 
However, since $\langle\tilde{\Phi}\rangle$, $\langle Y\rangle$
and $\langle \bar{E}\rangle$ can take diagonal forms 
simultaneously, those terms give the same contributions.
Therefore, for simplicity, we have denoted only the term 
 ${\rm Tr}[\tilde{\Phi}\tilde{\Phi}Y \bar{E}]$ as a typical
one of those in the expression (2.3).
(For the $\bar{\lambda}_Y$ term, the situation is similar
to the $\lambda_Y$ term.)
The potential $W_S$ plays a role in deriving the Sumino
relation.
Only in the potential $W_S$, we have assumed terms of new types
${\rm Tr}[AB]{\rm Tr}[CD]$ and  ${\rm Tr}[A]{\rm Tr}[B]
{\rm Tr}[C]{\rm Tr}[D]$  in addition to a type ${\rm Tr}[ABCD]$.
Since we have considered that those terms are written up
via two steps, a replacement 
${\rm Tr}[ABCD] \rightarrow {\rm Tr}[AB]{\rm Tr}[CD]$
and simultaneous replacements ${\rm Tr}[AB] \rightarrow 
{\rm Tr}[A]{\rm Tr}[B]$ and ${\rm Tr}[CD] \rightarrow
{\rm Tr}[C]{\rm Tr}[D]$, we did not consider such terms 
${\rm Tr}[ABC]{\rm Tr}[D]$, ${\rm Tr}[AB]{\rm Tr}[C]{\rm Tr}[D]$, 
and so on.

From the superpotential $W=W_Y +W_T +W_S$, we can obtain the 
following VEV relations. 
From a SUSY vacuum condition $\partial W/\partial \Theta_T=0$,
we obtain a VEV relation
$$
\bar{Y} = -\frac{\lambda_T}{\mu_T \Lambda} \bar{T} Y \bar{T} .
\eqno(2.5)
$$
A condition $\partial W/\partial \bar{T}=0$ leads to
$$
\frac{\partial W}{\partial \bar{T}} = \frac{\lambda_T}{\Lambda}
(Y \bar{T} \Theta_T +\Theta_T \bar{T} Y) =0 .
\eqno(2.6)
$$
Since we want $\langle Y\rangle \neq 0$ and 
$\langle \bar{T} \rangle \neq 0$, we choose a vacuum with
$\langle \Theta_T \rangle =0$. 
From SUSY vacuum conditions $\partial W/\partial \bar{Y}=0$
and $\partial W/\partial {Y}=0$,
we obtain
$$
Y\bar{E} = - \frac{\bar{\lambda}_Y}{\lambda_E} \Phi\Phi , 
\ \ \ E\bar{Y} = 
- \frac{{\lambda}_Y}{\lambda_E} \tilde{\Phi} \tilde{\Phi} .
\eqno(2.7)
$$
respectively, 
where we have used $\langle \Theta_T \rangle =0$.
Then, we can see that 
$\partial W/\partial E =0$ and $\partial W/\partial \bar{E} =0$ 
are automatically satisfied under the conditions (2.7).
For the moment, we assume the following VEV forms for $E$, 
$\bar{E}$ and $\bar{T}$:
$$
\frac{1}{v_E} \langle E\rangle = \frac{1}{\bar{v}_E} 
\langle \bar{E}\rangle = \left( 
\begin{array}{ccc}
1 & 0 & 0 \\
0 & 1 & 0 \\
0 & 0 & 1 
\end{array}
\right) , \ \ \  \frac{1}{\bar{v}_T} 
\langle \bar{T}\rangle = \left( 
\begin{array}{ccc}
0 & 0 & 1 \\
0 & 1 & 0 \\
1 & 0 & 0 
\end{array}
\right) , 
\eqno(2.8)
$$
on the basis in which $\langle\Phi\rangle$ takes a 
diagonal form $\langle\Phi\rangle = {\rm diag}
(v_1,v_2,v_3) \equiv v_0 {\rm diag}(z_1,z_2,z_3)$. 
(The forms of $\langle E\rangle$ and $\langle \bar{T} \rangle$ 
will be discussed in Sec.4.) 
Therefore, the forms $\langle Y\rangle$, $\langle \bar{Y}\rangle$
and $\langle \bar{\Phi}\rangle$ are expressed as
$\langle Y\rangle = v_Y Z^2$, $\langle \bar{Y}\rangle = 
\bar{v}_Y \bar{Z}^2$
and $\langle \tilde{\Phi}\rangle =\bar{v}_0 \bar{Z}$, where
$$
Z={\rm diag}(z_1,z_2,z_3), \ \ \ \bar{Z}={\rm diag}(z_3,z_2,z_1) ,
\eqno(2.9)
$$
$v_Y \bar{v}_E = -(\bar{\lambda}_Y/\lambda_E) v_0^2$,
$\bar{v}_Y {v}_E = -({\lambda}_Y/\lambda_E) \bar{v}_0^2$
and $\bar{v}_Y =-(\lambda_T/\mu_T \Lambda) v_Y$.

The Sumino relation is obtained as follows.
From a condition $\partial W/\partial \Phi=0$, we obtain
$$
\frac{\partial W}{\partial \Phi} = 
\frac{\bar{\lambda}_Y}{\Lambda}(\Phi E \bar{Y} +E \bar{Y} \Phi^T)
+\frac{\lambda_1}{\Lambda} (\Phi \tilde{\Phi} \tilde{\Phi} +
\tilde{\Phi} \tilde{\Phi} \Phi) +
2 \frac{\lambda'_1}{\Lambda} \tilde{\Phi} \Phi \tilde{\Phi} 
$$
$$  +
2\frac{\lambda_2}{\Lambda} {\rm Tr}[\tilde{\Phi} \tilde{\Phi}]\Phi
+2\frac{\lambda'_2}{\Lambda} {\rm Tr}[\Phi \tilde{\Phi}]\tilde{\Phi}
+ 2\frac{\lambda_3}{\Lambda} {\rm Tr}^2[\Phi]{\rm Tr}^2[\tilde{\Phi}]
{\bf 1}
=0 ,
\eqno(2.10)
$$
which leads to an equation for the parameters $z_i$
$$
c_0 {\bf 1}+ c_1 Z + c'_1 \bar{Z} + c_3 Z \bar{Z}^2 = 0 ,
\eqno(2.11)
$$
where $c_0 = \lambda_3 v_0 \bar{v}_0^2 (z_1+z_2+z_3)
 (z_1^2+z_2^2+z_3^2)$, 
$c_1=\lambda_2 v_0 \bar{v}_0^2 (z_1^2+z_2^2+z_3^2)$,
$c'_1=\lambda'_2 v_0 \bar{v}_0^2 (z_2^2+2 z_1 z_3)$,
and $c_3 = \bar{\lambda}_Y v_E \bar{v}_Y v_0 +
(\lambda_1+\lambda'_1)v_0 \bar{v}_0^2$.
Also, from $\partial W/\partial \tilde{\Phi} =0$,
we obtain
$$
\bar{c}_0{\bf 1} + \bar{c}_1 \bar{Z} +\bar{c}'_1 Z +
\bar{c}_3 \bar{Z} Z^2 = 0, 
\eqno(2.12)
$$
where $\bar{c}_0 = \lambda_3 \bar{v}_0 {v}_0^2 (z_1+z_2+z_3)
 (z_1^2+z_2^2+z_3^2)$, 
$\bar{c}_1=\lambda_2 \bar{v}_0 {v}_0^2 (z_1^2+z_2^2+z_3^2)$,
$\bar{c}'_1=\lambda'_2 \bar{v}_0 {v}_0^2 (z_2^2+2 z_1 z_3)$
and $\bar{c}_3 = {\lambda}_Y \bar{v}_E {v}_Y \bar{v}_0 +
(\lambda_1+\lambda'_1)\bar{v}_0 {v}_0^2$.
If we assume that $W_Y$ is symmetric under $\Phi\leftrightarrow
\tilde{\Phi}$, $Y \leftrightarrow \bar{Y}$ and 
$E \leftrightarrow \bar{E}$, so that 
$\bar{v}_0 =v_0$, $\bar{v}_Y=v_Y$ and $\bar{v}_E=v_E$, 
we find that Eq.(2.12) is effectively equivalent to Eq.(2.11).

Now, we solve Eq.(2.11), i.e. 
$c_0 + c_1 z_1+ c'_1 z_3 + c_3 z_3 z_1^2 = 0$,
$c_0 +(c_1+c'_1) z_2 + c_3 z_2^3 = 0$, and
$c_0 + c_1 z_3 + c'_1 z_1 + c_3 z_1 z_3^2 = 0$.
From the first and third equations, we obtain constraints on
the coefficients $c_0$, $c_1$, $c'_1$ and $c_3$, 
$$
c_1-c'_1 = c_3 z_1 z_3 , \ \ \ \ 
c_0 = -(c'_1 +c_3 z_1 z_3)(z_1+z_3) .
\eqno(2.13)
$$
By substituting Eqs.(2.13) into the second equation of (2.11),
we obtain $c_3 [z_2^3 + z_1 z_3 z_2 -z_1 z_3 (z_1+z_3)] + 
c'_1 (2 z_2-z_1-z_3)=0$, i.e.
$$
c_1[z_2^3 + z_1 z_3 z_2 -z_1 z_3 (z_1+z_3)] - c'_1(z_2^2-z_1 z_3) z_2 =0.
\eqno(2.14)
$$
Note that  Eq.(2.14) leads to the 
Sumino relation (1.5) in the limit of $c'_1\rightarrow 0$.
Since $c'_1/c_1$ is given by 
$$
\frac{c'_1}{c_1} = \frac{\lambda'_2}{\lambda_2} 
\frac{z_2^2 + 2z_1 z_3}{z_1^2+z_2^2+z_3^2} ,
\eqno(2.15)
$$
Eq.(2.14) is written as
$$
z_2^3+z_1 z_3 z_2 - z_1 z_3 (z_1+z_3) -\rho 
\frac{(z_2^2+2 z_1 z_3)(z_2^2-z_1 z_3)z_2}{z_1^2+z_2^2+z_3^2} = 0,
\eqno(2.16)
$$
where $\rho=\lambda'_2/\lambda_2$.

We can normalize the parameters $z_i$ as 
$$
z_1+z_2+z_3 = \sqrt{\frac{3}{2}} , \ \ \ 
z_1^2+z_2^2+z_3^2 = 1 + \xi ,
\eqno(2.17)
$$
without losing generality (by adjusting the value $v_0$
suitably). 
Then, the parameter $\xi$ corresponds to $\xi$ defined 
in Eq.(1.8).
Since $z_1+z_3 = \sqrt{3/2} -z_2$, and 
$$
z_1 z_3 = \frac{1}{2}[(z_1+z_3)^2-(z_1^2+z_3^2)]
=z_2^2-\sqrt{\frac{3}{2}} z_2 +\frac{1}{4} (1-2 \xi) ,
\eqno(2.18)
$$
we can regard Eq.(2.16) as an equation for $z_2$,
and we can solve it numerically. 
[The cubic equation (2.16) with $\rho=0$ (i.e. the Sumino
equation) has only one real solution, while a general
case with $\rho\neq 0$ has two real solutions.
Therefore, of two numerical solutions, we will choose 
a solution which is nearly equal to the Sumino solution.] 
For typical values of $\rho$, 
$\rho=(0, 1, 2)$, we obtain $\kappa =(2.20867, 2.1146,
 1.9792) \times 10^{-3}$ at $\xi=0$.
Therefore, the case with $\rho \simeq 1-2$ is in favor of the 
observed value $\kappa^{pole} =(2.0633\pm 0.001)\times 10^{-3}$,
if we assume that the Sumino mechanism correctly works.
However, at present, the parameter $\rho=\lambda'_2/\lambda_2$
is free, so that we can predict neither values of $\kappa(\mu)$
or $\xi(\mu)$. 
In the next section, we will speculate about the ratio 
$\lambda'_2/\lambda_2$.

\vspace{3mm}

\noindent{\large\bf 3 \ Speculation}

In this section, we speculate about a value of 
$\lambda'_2/\lambda_2$, and thereby, we speculate  
values of $\xi(\Lambda)$ and $\kappa(\Lambda)$.

In the previous section, we have considered that 
$\lambda_2$ and $\lambda'_2$ terms come from 
${\rm Tr}[ABCD]\rightarrow {\rm Tr}[AB] {\rm Tr}[CD]$.
Since ${\rm Tr}[AABB] = {\rm Tr}[BAAB] ={\rm Tr}[BBAA]
={\rm Tr}[ABBA]$ and ${\rm Tr}[ABAB] = {\rm Tr}[BABA]$,
we count the terms as 
${\rm Tr}[AABB] \rightarrow 2 ( {\rm Tr}[AA] {\rm Tr}[BB]
+{\rm Tr}[AB] {\rm Tr}[AB])$ and ${\rm Tr}[ABAB] \rightarrow
2 {\rm Tr}[AB] {\rm Tr}[AB]$, respectively.
Therefore, we consider
$$
 \lambda_1{\rm Tr}[\Phi\Phi\tilde{\Phi}\tilde{\Phi}]
+\lambda'_1{\rm Tr}[\Phi\tilde{\Phi}\Phi\tilde{\Phi}]
\rightarrow 
 \lambda_1{\rm Tr}[\Phi\Phi]{\rm Tr}[\tilde{\Phi}\tilde{\Phi}]
+ (\lambda_1+\lambda'_1){\rm Tr}[\Phi\tilde{\Phi}]
{\rm Tr}[\Phi\tilde{\Phi}] ,
\eqno(3.1)
$$
except for a common factor.
On the other hand, independent arrangements of two $A$ and two $B$
are $AABB$, $ABAB$, $ABBA$, $BBAA$, $BABA$ and $BAAB$,
so that we speculate $\lambda'_1/\lambda_1=1/2$.
Therefore, we obtain
$$
\rho \equiv \frac{\lambda'_2}{\lambda_2}=\frac{\lambda_1+
\lambda'_1}{\lambda_1} = \frac{3}{2} .
\eqno(3.2)
$$

The value $\rho=3/2$ predicts a value of $\kappa$
$$
\kappa(\xi=0) = 2.0653 \times 10^{-3} ,
\eqno(3.3)
$$
which is in excellent agreement with the observed 
value $\kappa^{pole}=(2.0633\pm 0.001) \times 10^{-3}$.
However, as we have shown in Table 1, 
in the yukawaon model, the value
of $\xi$ will be not zero, unless 
the Sumino mechanism exactly works.

In Fig.2, we illustrate a predicted   
$\xi(\mu)$-$\kappa(\mu)$ relation, where the predicted 
values $\kappa(\mu)$ are estimated for input values of
$\xi(\mu)$ which correspond to the values of 
$\xi^{obs}(\mu)$ at $\mu=m_Z$, $10^3$ GeV, $10^9$ GeV, 
$10^{12}$ GeV and $2\times 10^{16}$ GeV in Table 1, 
respectively.
As seen in Fig.2, the curve of $\xi(\mu)$-$\kappa(\mu)$ is
crossed with the curve of $\xi(\mu)^{obs}$-$\kappa^{obs}(\mu)$
at $\mu \sim 10^4$ GeV.
Of course, the crossed point depends on the value of $\tan\beta$. 
For a reference, in Fig.3, we also plot a $\xi$-$\kappa$ 
relation for the case with $\tan\beta=50$.
The curves for $\rho=3/2$ and $\rho=1$ have crossing points
with the $\xi^{obs}$-$\kappa^{obs}$ curve at $\mu \sim 10^2$ GeV 
and $\mu \sim 10^{11}$ GeV, respectively. 
If we accept the speculated value $\rho=3/2$, we are obliged
to consider $\Lambda \sim 10^2$ GeV, so that the case with
$\tan\beta=50$ is unlikely.
We consider that the value of $\Lambda$ is of the order of
$10^4$ GeV or more, so that the value of $\tan\beta$ must be 
of the order of $10$ or less.


\vspace{3mm}

\noindent{\large\bf 4 \ VEV matrix forms 
$\langle E\rangle$ and $\langle \bar{T}\rangle$}

So far, we have not mentioned the origin of the VEV forms
$\langle E\rangle$ and $\langle \bar{T}\rangle$.
In Sec.2, we have a priori assumed their VEV forms (2.8).
In a non-SUSY model, we can readily give an explicit 
scalar potential form which leads to the form
$\langle E\rangle = v_E {\bf 1}$. 
However, in general, in a SUSY model, it is hard to give
a superpotential which leads to the form
$\langle E\rangle = v_E {\bf 1}$, although we can give 
$\langle E\rangle \langle \bar{E} \rangle \propto {\bf 1}$.
For example, we assume that a super potential
$$
W_E = \frac{\lambda_1^E}{\Lambda} {\rm Tr}[E\bar{E}E\bar{E}]
+\frac{\lambda_2^E}{\Lambda} {\rm Tr}^2[E\bar{E}] ,
\eqno(4.1)
$$
where we have chosen $R$ charge parameters as 
$Q_R(E\bar{E})=r_1+2 r_2=1$.  
Supersymmetric vacuum conditions 
$\partial W_E/\partial E =0$ and 
$\partial W_E/\partial \bar{E} =0$ demand 
$$
\lambda_1^E \bar{E} E + \lambda_2^E {\rm Tr}[\bar{E} E]
{\bf 1} =0 ,
\eqno(4.2)
$$
together with an additional condition for the coefficients
${\lambda_2^E}/{\lambda_1^E} = -{1}/{3}$.
%
One way to obtain a solution $\langle E\rangle = v_E {\bf 1}$
from $\bar{E} E \propto {\bf 1}$ 
is to assume that our U(3) flavor symmetry is
broken into O(3) above $\mu\sim\Lambda$ by 
$\langle E\rangle$, 
because we know that a U(3) symmetry is broken into 
an O(3) symmetry
when a field ${\bf 6}$ of U(3) takes a VEV form 
$\langle {\bf 6}\rangle \propto {\bf 1}$. 

Another way is to assume that our U(3) is gauge symmetry
according to Sumino's idea \cite{Sumino09PLB,Sumino09JHEP}.
Then, we can use a $D$-term condition for fields ${\bf 6}$
and $\bar{{\bf 6}}$ which are 6-plet and $6^*$-plet of U(3),
respectively:
$$
\sum_{\bf 6} {\rm Tr} [{\bf 6}^\dagger \lambda_a {\bf 6}
+ {\bf 6}\lambda_a^T {\bf 6}^\dagger ] -
\sum_{\bar{{\bf 6}}} {\rm Tr} [\bar{\bf 6}^\dagger 
\lambda_a \bar{\bf 6}
+ \bar{\bf 6}\lambda_a^T \bar{\bf 6}^\dagger ] =0 ,
\eqno(4.3)
$$
where $\lambda_a$ ($a=1,2,\cdots,8$) are Gell-Mann's 
$\lambda$ matrices in SU(3) and $\lambda_0=\sqrt{2/3}
\,{\bf 1}$.
(The $D$-term condition does not put any constraint on 
the fields $\Phi$ and $\tilde{\Phi}$ which are 
${\bf 8}+{\bf 1}$ of U(3).)
If we assume that $\langle {\bf 6}\rangle$ and 
$\langle \bar{{\bf 6}}\rangle$ are diagonal, 
the condition (4.3) is applied only to the cases 
$\lambda_3$, $\lambda_8$ and $\lambda_0$. 
A solution $E=\bar{E}=v_E {\bf 1}$ under the constraint 
$E \bar{E} \propto {\bf 1}$ satisfies a relation
$$
{\rm Tr} [ E^\dagger \lambda_a E+ E\lambda_a^T E^\dagger] 
- {\rm Tr} [\bar{E}^\dagger \lambda_a \bar{E}
+ \bar{E}\lambda_a^T \bar{E}^\dagger ] =0 .
\eqno(4.4)
$$
Similarly, if we introduce a field $T$, which is ${\bf 6}$ 
of U(3), in addition to $\bar{T}$, we can obtain 
$\langle T\rangle \langle \bar{T}\rangle \propto {\bf 1}$.
Considering $\langle T\rangle^T = \langle T\rangle$ and 
$\langle \bar{T}\rangle^T = \langle \bar{T}\rangle$,
we take specific forms 
$$
\langle T \rangle = \left(
\begin{array}{ccc}
0 & 0 & v_1 \\
0 & v_2 & 0 \\
v_1 & 0 & 0 
\end{array} \right) , \ \ \ \ 
\langle \bar{T} \rangle = \left(
\begin{array}{ccc}
0 & 0 & \bar{v}_1 \\
0 & \bar{v}_2 & 0 \\
\bar{v}_1 & 0 & 0 
\end{array} \right) ,
\eqno(4.5)
$$
of solutions for 
$\langle T\rangle \langle \bar{T}\rangle \propto {\bf 1}$.
Then, we can also see that a solution 
$\langle {T} \rangle =\langle \bar{T} \rangle$ with 
$v_1=v_2$ under the constraint $T \bar{T} \propto {\bf 1}$
satisfies 
$$
{\rm Tr} [ T^\dagger \lambda_a T+ T\lambda_a^T T^\dagger] 
- {\rm Tr} [\bar{T}^\dagger \lambda_a \bar{T}
+ \bar{T}\lambda_a^T \bar{T}^\dagger ] =0 .
\eqno(4.6)
$$
Of the fields with ${\bf 6}$ and $\bar{\bf 6}$ of U(3) in
the present model, for the last one $\Theta_T$, 
the VEV has to be $\langle \Theta_T \rangle =0$ from the
$D$-term condition (4.3), because we already have 
relations (4.4) and (4.6) for $E$, $\bar{E}$, $T$
and $\bar{T}$. 
This solution $\langle \Theta_T \rangle =0$ is consistent
with the requirement $\langle \Theta_T \rangle =0$ in
Eq.(2.6).



\vspace{3mm}

\noindent{\large\bf 5 \ Concluding remarks}

In conclusion, on the basis of a yukawaon model, we have 
investigated  the Sumino relation  (1.2) for the 
charged lepton masses.
As well as the Sumino model \cite{Sumino09JHEP}, 
where there is a $\lambda_2^S$ term in addition 
to the Sumino term (the $\lambda_1^S$ term)
in the scalar potential (1.4), in the present model, too,
we have additional terms, $c'_1$ terms, in addition to
the Sumino terms ($c_1$-terms) in Eq.(2.14).
In the Sumino model, since the contribution of the 
$\lambda^S_2$ term is considerably large compared with 
that of the $\lambda^S_1$ term for the realistic mass 
values, we must consider that $\lambda_2^S/\lambda_1^S
\sim 10^{-2}$ in order to give the observed electron
mass value. 
In other words, a predicted value of the electron mass 
is highly sensitive to the second term ($\lambda_2^S$-term).
On the other hand, in the present model, 
the $c'_1$ terms less contribute to 
the essential terms ($c_1$ terms), so that existence 
of the $c'_1$ terms does not so spoil the Sumino equation
(1.5) even if we take $\rho \sim 1$, and rather, 
the existence is in favor of the numerical fitting.   

We would like to emphasize that we have never used 
the charged lepton mass relation (1.1). 
Although we have put $K=(2/3)(1+\xi)$, we have never
assumed that the parameter $\xi$ is small.
Nevertheless, as shown in Figs.2 and 3, we can conclude
that the values of $\xi(\Lambda)$ must be considerably 
small in order to accommodate the predicted $\xi$-$\kappa$
curves to the running mass values.  

If we accept a speculation $\lambda'_2/\lambda_2
=3/2$, we obtain two options:
One option is to consider that we accept the Sumino mechanism,
so that the prediction (3.3) at $\xi=0$ is excellently in 
favor of the observed value (1.7) of $\kappa^{pole}$.
Another option is to consider that the energy scale $\Lambda$ 
in the yukawaon model is $\Lambda \sim 10^4$ GeV
because the curve of the predicted value $\kappa(\mu)$ 
crosses with the curve of the observed value $\kappa(\mu)^{obs}$
at $\mu \sim 10^4$ GeV as seen in Fig.2. 
(We have assumed that U(3) is not gauged, or that the contributions
due to U(3) gauge bosons are negligibly small.) 
The latter scenario provides fruitful phenomenology in TeV
region physics.
Since in past yukawaon models the energy scale $\Lambda$ was 
taken as $\Lambda \sim 10^{14-16}$ GeV, all results which 
were obtained 
from the old yukawaon models must be re-considered.

In the present model, the VEV of the yukawaon $Y_e$ is given
by $\langle Y_e \rangle \propto  \langle \Phi\rangle
\langle E_e\rangle \langle \Phi\rangle^T$ (exactly speaking, 
 $\langle Y_e \rangle \propto \lambda_{Y1} \langle \Phi\rangle
\langle E_e\rangle \langle \Phi\rangle^T + \lambda_{Y2} (
\langle \Phi\rangle \langle \Phi\rangle \langle E_e \rangle
+\langle E_e \rangle\langle \Phi\rangle^T \langle \Phi\rangle^T)$).
Note that $\Phi$ (and $\bar{\Phi}$) have no sector charge, and 
only the filed $E_e$ has a non-zero sector charge $Q_X$.
Therefore, we may consider a general form
$$
\langle Y_f \rangle \propto  \langle \Phi\rangle
\langle E_f\rangle \langle \Phi\rangle^T .
\eqno(5.1)
$$
Since we have assumed that $\langle E_e \rangle$ is real,  
we could reduce 
${\rm Tr}[Y_e]$ to ${\rm Tr}[D_e] \propto m_e + m_\mu +m_\tau$ 
($D_f$ is a diagonalized matrix defined by
 $U_f^T Y_f U_f =D_f$), while, for a general yukawaon $Y_f$,
which is ${\bf 6}$ of U(3), i.e. which is not Hermitian, 
we cannot reduce ${\rm Tr}[Y_f]$ to such a form, because 
$U_f U_f^T \neq {\bf 1}$.
Therefore, only for the charged lepton sector, we can express
the quantities $K$ and $\kappa$   
by the ratios ${\rm Tr}[\Phi \Phi^T]/{\rm Tr}^2[\Phi]$
and ${\rm det}(\Phi)/{\rm Tr}^3[\Phi]$, respectively, 
while, for other sectors, 
we cannot obtain such relations (1.3) and (1.6).
We think that this is a reason that $K_f \simeq 2/3$ are not 
satisfied \cite{dev-K} in other sectors (e.g. quark sectors).  

As we have stated in Sec.1, although the present topic 
appears to be narrow and restricted, we think that the 
problems proposed by Sumino will provide a vital clue 
to new physics.
If we take the speculation $\Lambda \sim 10^4$ GeV 
seriously, we may expect possibility that the model
provides fruitful and visible effects in TeV region 
physics.
We hope that the present model can give some hints for
fruitful developments of the Sumino model.

\vspace{3mm}

\centerline{\large\bf Acknowledgments}

The author would like to thank Y.~Sumino for valuable and 
helpful conversations (and also Particle Physics Group
at Tohoku University for their hospitality). 
He also thanks S.~Yamaguchi, T.~Yamashita, Y.~Hyakutake
and N.~Uekusa for helpful discussions and comments on 
a framework of a SUSY scenario. 
Especially, he is indebted to Yamashita for the idea of
$D$-term condition.
This work is supported by the Grant-in-Aid for
Scientific Research (C), JSPS, (No.21540266).


\newpage

{\scalebox{0.7}{\includegraphics{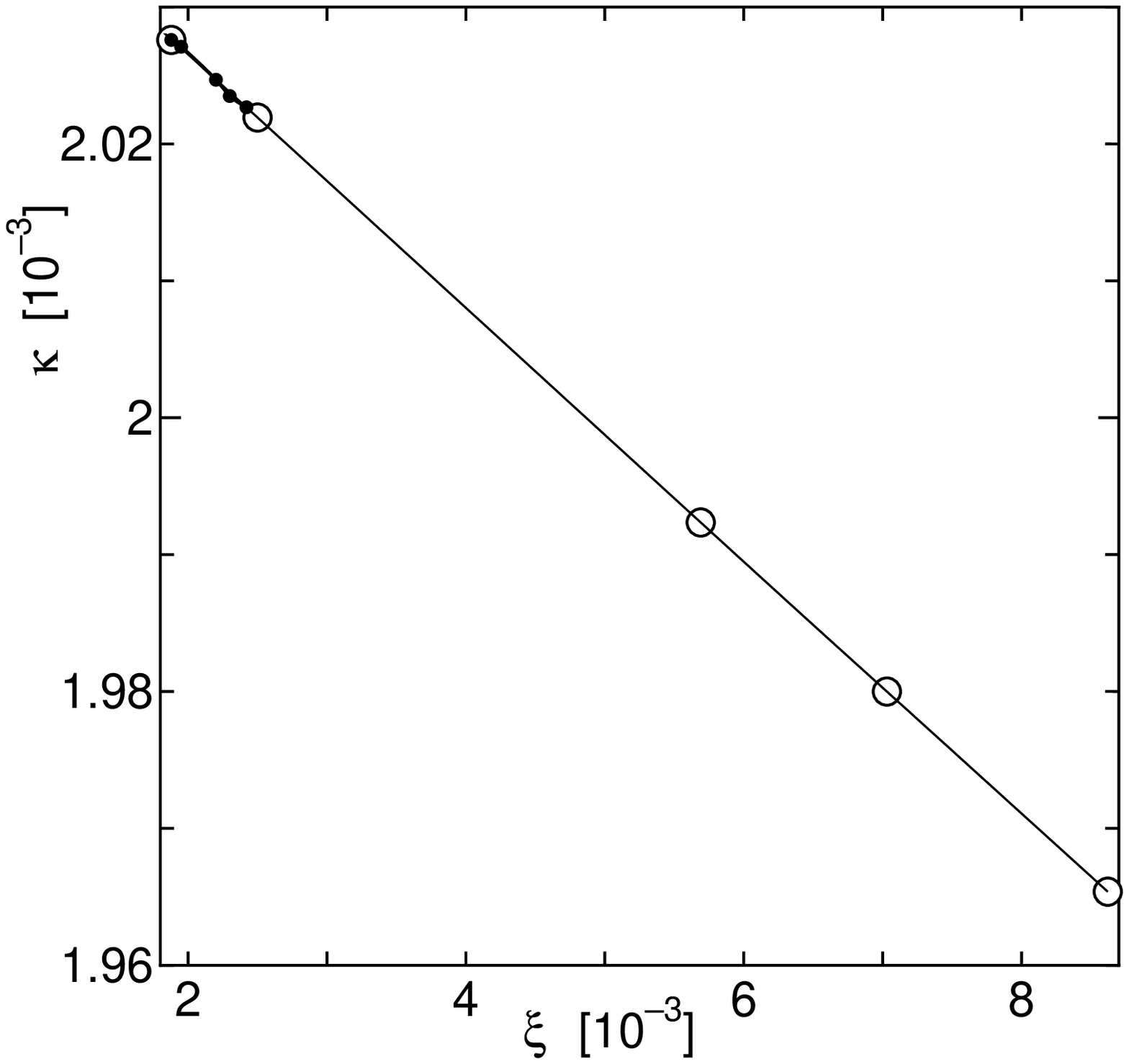}} }

\begin{quotation}
{\bf Fig.~1}  
Relation between $\xi(\mu)$ and $\kappa(\mu)$ estimated from
running masses.  One (a thick line with filled-in circle points) is 
for the case $\tan\beta=10$, another one (a thin line with opened 
circle points) is for $\tan\beta=50$. 
The data have been quoted from Xing {\it et al.} \cite{Xing08}.
The data points from the left to the right
correspond to $\xi$-$\kappa$ values at $\mu=m_Z$, $10^3$ GeV, 
$10^9$ GeV, $10^{12}$ GeV and $2\times 10^{16}$ GeV, respectively.
\end{quotation}

\newpage

{\scalebox{0.7}{\includegraphics{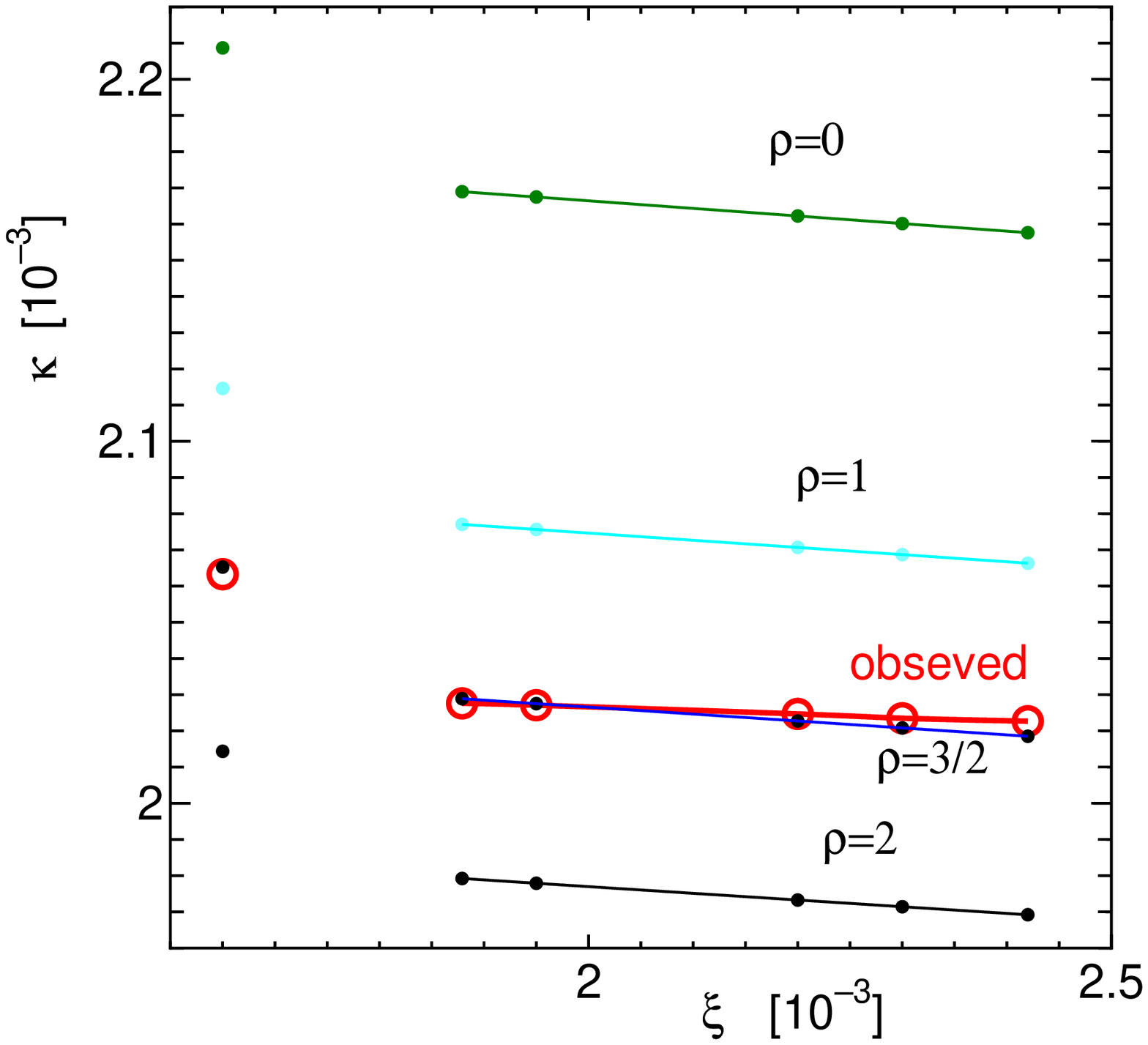}} }

\begin{quotation}
{\bf Fig.~2}  
Relation between $\xi(\mu)$ and $\kappa(\mu)$ for the case
$\tan\beta=10$.
A thick line with opened circles is plotted by values 
given in Table 1.
A curve for $\rho=0$ denotes the Sumino relation (1.2).
Curves for $\rho=1$, $3/2$ and $2$ are examples of 
predictions in the present model.   
The data points from the left to the right correspond to  
$\xi$-$\kappa$ values at $\mu=m_Z$, $10^3$ GeV, $10^9$ GeV,
$10^{12}$ GeV and $2\times 10^{16}$ GeV, respectively.
On the left of those $\xi$-$\kappa$ curves, for a reference,
values of $\kappa$ at $\xi=0$ are plotted together with 
observed value $\kappa^{pole}$. 
\end{quotation}

\newpage

{\scalebox{0.7}{\includegraphics{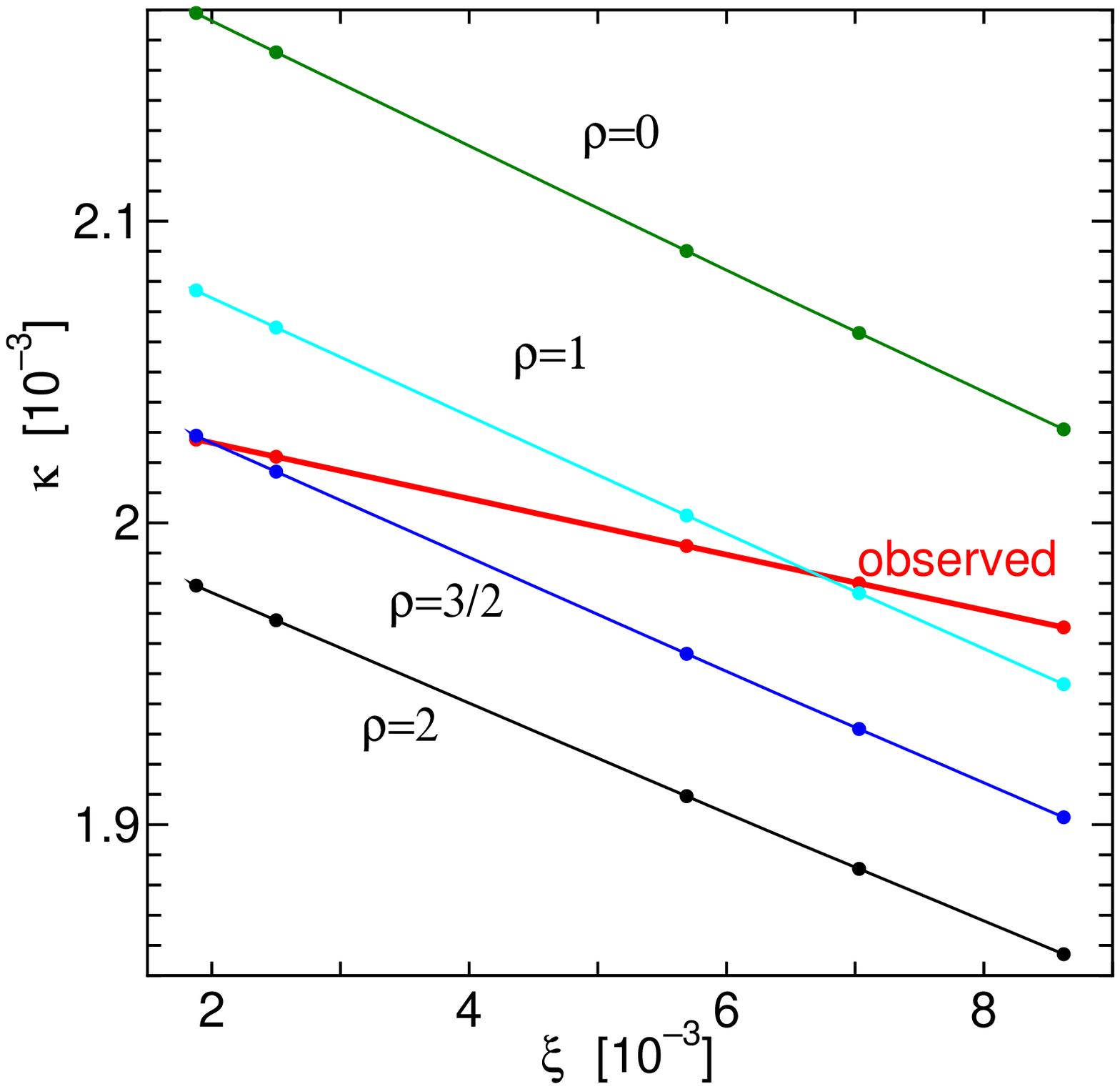}} }

\begin{quotation}
{\bf Fig.~3}  
Relation between $\xi(\mu)$ and $\kappa(\mu)$ for the case
$\tan\beta=50$.
The data points from the left to the right correspond to  
$\xi$-$\kappa$ values at $\mu=m_Z$, $10^3$ GeV, $10^9$ GeV,
$10^{12}$ GeV and $2\times 10^{16}$ GeV, respectively.
\end{quotation}

\end{document}